\documentclass{article}
\usepackage[T1]{fontenc} % add special characters (e.g., umlaute)
\usepackage[utf8]{inputenc} % set utf-8 as default input encoding
\usepackage{ismir,amsmath,cite,url}
\usepackage{graphicx}
\usepackage{color}
\usepackage{booktabs}
\usepackage{subcaption}
\usepackage{algpseudocode}
\usepackage{algorithm}
\usepackage[bookmarks=false]{hyperref}

\newcommand\rurl[1]{%
  \href{https://#1}{\nolinkurl{#1}}%
}

\newcommand{\var}{}

\newcommand{\Parameter}[2]{\Statex $\triangleright$ \var{#1}: #2}

\newcommand{\dataset}{\texttt{MidiCaps}}

\usepackage{lineno}
\title{MidiCaps: A large-scale MIDI dataset with text captions}

\oneauthor
    {Jan Melechovsky $^*$, Abhinaba Roy $^*$, Dorien Herremans}
    {Singapore University of Technology and Design \\ jan\_melechovsky@mymail.sutd.edu.sg, abhinaba\_roy@sutd.edu.sg, dorien\_herremans@sutd.edu.sg}

\def\authorname{J. Melechovsky, A. Roy, and D. Herremans}

\begin{document}
% \nolinenumbers
%
\maketitle
\begin{abstract}
% Generative models guided by text prompts are increasingly becoming more popular. However, no text-to-MIDI models currently exist, mostly due to the lack of a captioned MIDI dataset. This work aims to enable research that combines LLMs with symbolic music by presenting the first large-scale MIDI dataset with text captions that is openly available: \textbf{\dataset{}}. MIDI (Musical Instrument Digital Interface) files are a widely used format for encoding musical information. Their structured format captures the nuances of musical composition and has practical applications by music producers, composers, musicologists, as well as performers. Inspired by recent advancements in captioning techniques applied to various domains, we present a large-scale curated dataset of over 168k MIDI files accompanied by textual descriptions. Each MIDI caption succinctly describes the musical content, encompassing tempo, chord progression, time signature, instruments present, genre and mood; thereby facilitating multi-modal exploration and analysis. The dataset contains a mix of various genres, styles, and complexities, offering a rich source for training and evaluating models for tasks such as music information retrieval, music understanding and cross-modal translation. We provide detailed statistics about the dataset and have assessed the quality of the captions in an extensive listening study. 
% We anticipate that this resource will stimulate further research in the intersection of music and natural language processing, fostering advancements in both fields.
Generative models guided by text prompts are increasingly becoming more popular. However, no text-to-MIDI models currently exist due to the lack of a captioned MIDI dataset. This work aims to enable research that combines LLMs with symbolic music by presenting \textbf{\dataset{}}, the first openly available large-scale MIDI dataset with text captions. MIDI (Musical Instrument Digital Interface) files are widely used for encoding musical information and can capture the nuances of musical composition. They are widely used by music producers, composers, musicologists, and performers alike. Inspired by recent advancements in captioning techniques, we present a curated dataset of over 168k MIDI files with textual descriptions. Each MIDI caption describes the musical content, including tempo, chord progression, time signature, instruments, genre, and mood, thus facilitating multi-modal exploration and analysis. The dataset encompasses various genres, styles, and complexities, offering a rich data source for training and evaluating models for tasks such as music information retrieval, music understanding, and cross-modal translation. We provide detailed statistics about the dataset and have assessed the quality of the captions in an extensive listening study. We anticipate that this resource will stimulate further research at the intersection of music and natural language processing, fostering advancements in both fields.
% Additionally, we provide baseline experiments and evaluations to showcase the utility and potential applications of the \dataset{} dataset.
\end{abstract}
\def\thefootnote{*}\footnotetext{These authors contributed equally to this work.}\def\thefootnote{\arabic{footnote}}

% Points to mention in abstract: 

% - First text caption annotated dataset
% - we will leverage LLMs and years of MIR research to create this dataset
% - openly released (creative commons) dataset
% - Validated with user study

%
\section{Introduction}
\label{sec:introduction}
The recent development of large-language models (LLMs) has revolutionised how we interact with text, images, and even audio. By incorporating elements of multimodal learning, researchers have combined LLMs with other modalities. The resulting models can analyze and generate accurate descriptions and captions, which in turn facilitates downstream tasks such as question answering~\cite{touvron2023llama}, image generation~\cite{ramesh2021zero}, and music generation~\cite{melechovsky2023mustango}. However, we have yet to see such an evolution for MIDI files. 
\begin{figure}[t]
\includegraphics[width=8cm]{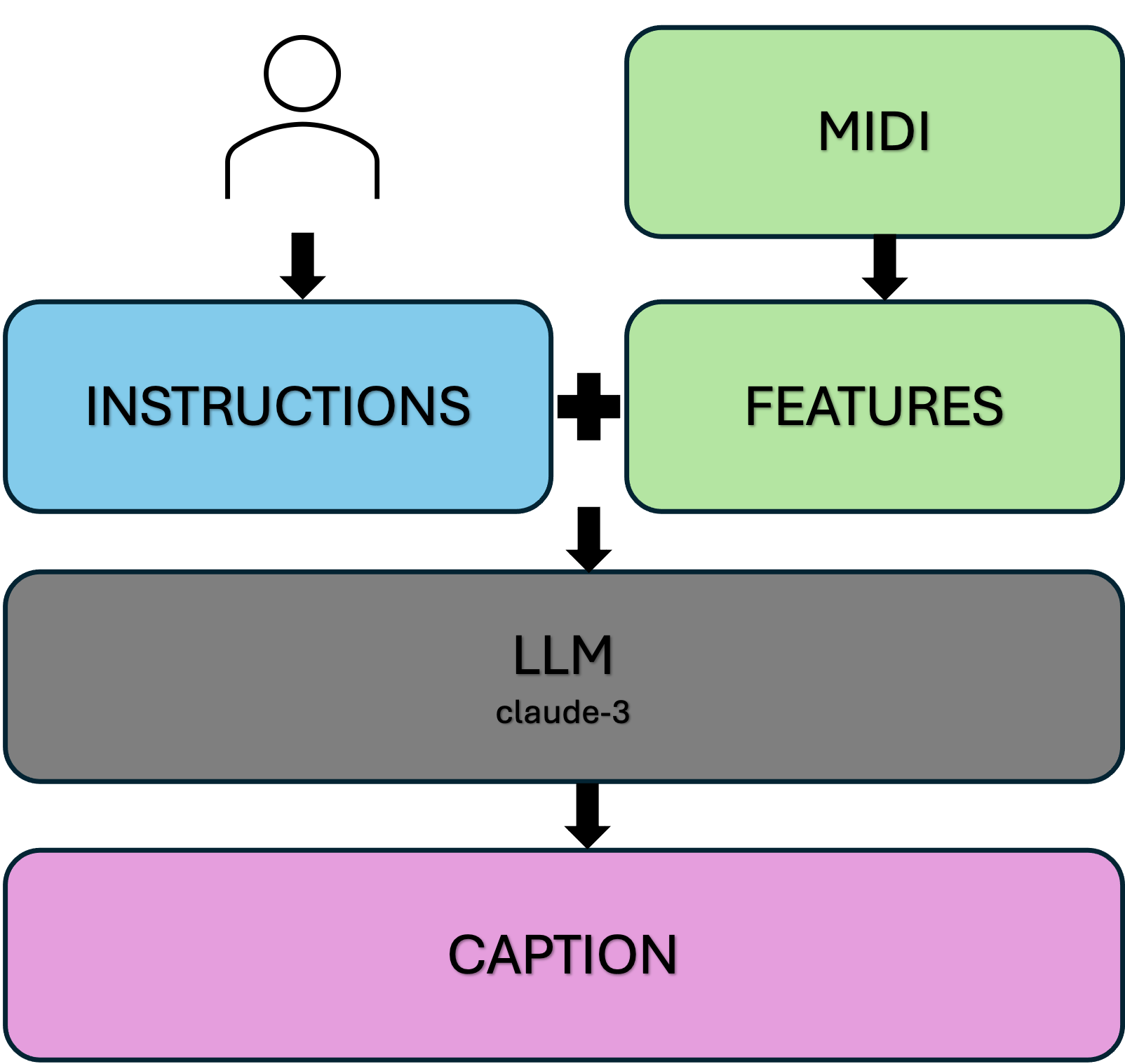}
\caption{Overview of our approach. We extract meaningful and relevant features from MIDI files. These features are then added as tags to the human instructions that are sent to an LLM (Claude-3) to generate meaningful text captions of MIDI files. }
\label{fig:overview}
\vspace{-4mm}
\end{figure}
%A stepping stone for harnessing the power of LLMs is a large dataset with proper texual descriptions that can serve as a benchmarking tool for target tasks such as  This may be due to a lack of text annotated MIDI datasets to train large-scale models. In this paper, we address this challenge by presenting the first large-scale MIDI dataset with text annotations that describe the musical content. 

% MIDI (Musical Instrument Digital Interface) is a standardized protocol that facilitates communication between electronic musical instruments, computers, and other devices. It represents musical events such as note-on, note-off, pitch, velocity, and control changes in a digital format, enabling precise control and manipulation of music-related parameters. 
In the field of Music Information Retrieval (MIR), MIDI plays a crucial role as a symbolic and musically meaningful representation of music. The format is often used by music producers and composers working in Digital Audio Workstations (DAWs). It is also a useful format for the computational analysis of music and related tasks such as music transcription, genre classification, similarity measurement, and music recommendation \cite{schedl2014music}. Furthermore, due to the symbolic nature of music, it has long been used by music generation algorithms \cite{herremans2017functional}.
% facilitating the creation of new compositions based on existing patterns or styles.
In recent years, we have seen a surge of interest in music generation from free-flow text instructions \cite{agostinelli2023musiclm,huang2023noise2music,melechovsky2023mustango,liu2023audioldm2,copet2023simple}. These studies leverage the expressive capabilities of LLMs to translate textual representation of musical attributes into actual music audio. This necessitates a meticulous alignment between the textual and musical feature spaces to ensure that the generated music closely follows text instructions. To validate and establish benchmarks for this text-to-music mapping, large-scale datasets with text captions have been developed \cite{melechovsky2023mustango,doh2023lp}.\par
No such efforts, however, have yet been made for the MIDI format, despite its widespread use by musicians and its obvious, historically supported usage in music generation. This lack of text-MIDI datasets, in turn, has inhibited researchers from exploring interesting and novel tasks such as MIDI generation from free-flow text prompts. In this work, we identify this shortcoming and develop a robust solution in the form of a large-scale curated MIDI dataset accompanied by text captions. Our goal is to obtain captions that are i) large in volume, ii) contain accurate information about the musical contents, and iii) feature a rich and refined vocabulary. We posit that such a dataset-level approach opens up further opportunities for researchers in MIDI-LLMs-related tasks.\par
To address the first goal, we identify an open source large-scale MIDI dataset in the form of Lakh MIDI dataset \cite{raffel2016learning}, that contains over 170K MIDI examples. Second, to attain the musical contents in each MIDI file, we extract meaningful features encompassing tempo, chord progression, time signature, instruments present, genre and mood. Each of the features are extracted using state-of-the-art MIR tools that ensure the quality and accuracy of the features extracted. After feature extraction, we are still left with the task of caption generation. Relying on traditional human annotation is tedious, time-consuming, and costly. Instead, motivated by the recent success of LLMs, we utilize in-context learning -- a model's ability to temporarily learn from human-provided instructions \cite{wei2022emergent}. Our decision is motivated by Melechovsky et al. \cite{melechovsky2023mustango}, who have demonstrated the efficacy of in-context learning in generating captions that are accurate, rich in description as well as grammatically coherent. In our approach, we furnish the LLM with instructions to generate captions based on the extracted music features, supplemented by a small set of feature-caption pairs created by expert annotators. Given the current absence of freely available MIDI-caption datasets, we anticipate that the provision of a substantial volume of detailed and informative captions will inspire the research community to delve further into tasks related to MIDI and Large Language Models (LLMs).
The main contributions of this work can be summarized as follows:
    % 1) We introduce the first curated large-scale open dataset of MIDI-caption pairs, termed \textbf{\dataset}\footnote{\rurl{huggingface.co/datasets/amaai-lab/MidiCaps} };
    % 2) We present a comprehensive set of music-specific features extracted from MIDI files. These features succinctly characterize the musical content, encompassing tempo, chord progression, time signature, instrument presence, genre, and mood;
    % 3) We provide a text caption annotation framework tailored specifically for MIDI data (see Figure~\ref{fig:overview}). Leveraging the in-context learning capability of large language models (LLMs), we enable the generation of captions using only a small number of feature-caption training pairs. This framework, a first of its kind, is made freely accessible to users\footnote{\rurl{github.com/AMAAI-Lab/MidiCaps}}, facilitating the generation of MIDI-caption pairs for their individual MIDI files.
\begin{itemize}
    \setlength\itemsep{0em}
    \item We introduce the first curated large-scale open dataset of MIDI-caption pairs, termed \textbf{\dataset}\footnote{\rurl{huggingface.co/datasets/amaai-lab/MidiCaps} }.
    % \vspace{-4mm}
    \item Furthermore, we present a comprehensive set of music-specific features extracted from MIDI files. These features succinctly characterize the musical content, encompassing tempo, chord progression, time signature, instrument presence, genre, and mood.
    % \vspace{-4mm}
    \item Finally, we provide a text caption annotation framework tailored specifically for MIDI data (see Figure~\ref{fig:overview}). Leveraging the in-context learning capability of large language models (LLMs), we enable the generation of captions using only a small number of feature-caption training pairs. This framework, a first of its kind, is made freely accessible to users\footnote{\rurl{github.com/AMAAI-Lab/MidiCaps}}, facilitating the generation of MIDI-caption pairs for their individual MIDI files. 
\end{itemize}

% \dh{Talk about motivation, also that this may lead to eventual text-to-music systems like Mustango, but then for midi. }
% \jm{yesyes, motivation, and inspiration from a recent paper Mustango where they do musicbench!!!}

\section{Related Work}\label{sec:relwork}

To the best of our knowledge, there are no publicly available MIDI caption datasets. In this section, we briefly mention various publicly available MIDI datasets and discuss the closely related topic of caption generation from audio and music.

Despite the scarcity of MIDI caption datasets, existing repositories offer potential resources that could be adapted for this purpose. Among these, the Lakh MIDI Dataset \cite{raffel2016learning} stands out, comprising a vast collection of MIDI files. While primarily tailored for MIR tasks such as melody extraction and chord estimation, its volume and diversity present an opportunity for repurposing towards captioning tasks, albeit requiring appropriate preprocessing. The MAESTRO Dataset \cite{hawthorne2018enabling} offers aligned pairs of MIDI and audio files, primarily for piano music generation. The MuseGAN Dataset \cite{dong2018musegan} focuses on multi-track songs, and the MAPS Dataset \cite{ycart2018maps}, contains recordings of classical piano pieces alongside aligned MIDI files and thus also present potential avenues for MIDI captioning research. Additionally, the Wikifonia Dataset \cite{simonetta2018symbolic} features a substantial collection of lead sheets accompanied by MIDI files. Closest to our proposed MIDI-caption dataset is the WikiMusicText (WikiMT) dataset \cite{wu2023clamp}, which includes lead sheets in ABC notation with metadata including text descriptions. These descriptions, however, pertain to general information about the music piece rather than detailed descriptions of musical contents provided in MIDI files within our captions.

In the last three years, several models were released for automatic caption generation from music \texttt{audio} files. One of the earlier models, MusCaps \cite{manco2021muscaps}, uses an architecture based on recurrent and convolutional layers as well as a multimodal encoder. Recent research includes the use of large language models (LLMs) for captioning \cite{doh2023lp,melechovsky2023mustango,huang2023noise2music}.
% In \cite{huang2023noise2music}, a pseudo labeling method is used to label a large training dataset. First, existing captions from other datasets are curated, then the MuLaN \cite{DBLP:conf/ismir/HuangJLGLE22} model, a joint music-text embedding model, evaluates the distance between captions and unlabeled audio files. Top caption candidates are selected based on their frequency to ensure balance among all samples.
In \cite{huang2023noise2music}, a pseudo labeling approach is used to label a large training dataset. First, existing captions from other datasets are curated, then the MuLaN \cite{DBLP:conf/ismir/HuangJLGLE22} model, a joint music-text embedding model, evaluates the distance between captions and unlabeled audio files. The top caption candidates are selected based on their frequency to ensure balance among all samples.
In \cite{kuang2022music}, the focus is on capturing the full sentiment of classical music recordings through text descriptions, introducing a Group-Topology Preservation Loss to be used with their cross-modality translation model. A recent study by Doh et al. \cite{doh2023lp} targets pseudo labeling of audio data with the help of an LLM, utilizing the MusicCaps \cite{agostinelli2023musiclm} dataset as ground truth and instructing GPT-3.5 Turbo \cite{ouyang2022training} to generate full captions from these tags.

In \cite{melechovsky2023mustango}, Melechovsky et al. curate a new dataset based on the MusicCaps dataset \cite{agostinelli2023musiclm}, called MusicBench. In MusicBench, the original captions are enhanced by including additional music descriptors such as chord sequence, musical key, time signature, and tempo. After performing audio and text augmentations to expand the dataset size, they use ChatGPT \cite{chatgpt} for rephrasing captions to create more diverse captions. Furthermore, they employ in-context learning to guide ChatGPT using a small set of human-annotated examples, instructing it to generate diverse captions to create an evaluation dataset from extracted tags, named FMACaps. Inspired by their methodology, we adopt a similar approach and utilize in-context learning alongside a large-language model to generate captions from MIDI features. In the subsequent section, we offer an in-depth description of our proposed framework for MIDI captioning.

% \dh{Few sentences about what to transition about what we are doing and then details described in next section. }

% % There is no related -- there is: wikiMusicText, and audio captioning, plus datasets on midi etc. 

% % \dh{You can list existing midi datasets (based on Nima's list. Maybe even in a table. Then you can talk about similar work in Audio/music to text see table here }\url{https://docs.google.com/document/d/1gAz9BY0cNN-OqQMlQJPEHAi4orJuaH3MlfkgcFb8Y9E/edit?usp=sharing})

% % CLAMP: \cite{wu2023clamp}

% Another multimodal dataset for midi (not text): \cite{cardoso2024nes} NES Video-Music Database: A Dataset of Symbolic Video Game Music Paired with Gameplay Videos.  \dh{Can comment if no time to write up}

\begin{figure*}[h!]
\centering
\includegraphics[width=9cm]{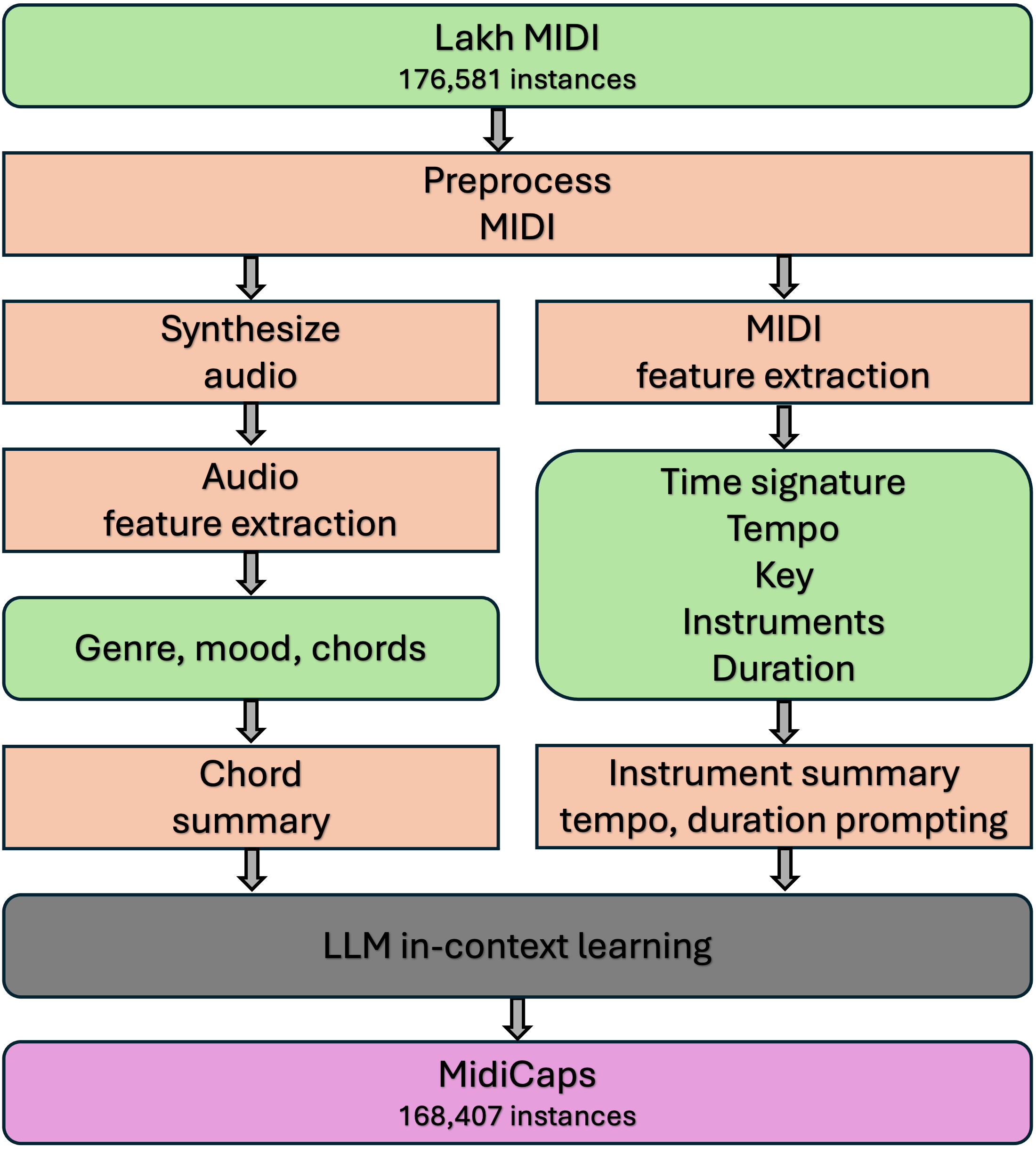}
\caption{Detailed overview of our proposed captioning framework.}
\label{fig:detailed_overview}
\vspace{-4mm}
\end{figure*}

\section{Method}
In this section, we discuss details regarding the music-specific features we extract from MIDI files.

\subsection{Feature extraction}
\label{sec:feat}
In a first step, as per Figure~\ref{fig:detailed_overview}, we extract various musical features from the MIDI files. This is achieved in two  ways: a number of features are extracted directly from the MIDI files, and others are extracted from the synthesized MIDI files. The details of our approach are described below. 

\subsubsection{Preprocessing}
We preprocess all files to remove faulty files. For instance, we found multiple files that had never-ending notes. Using Mido \cite{mido}, we further exclude files of duration shorter than 3 seconds and longer than 15 minutes.

\subsubsection{MIDI feature extraction}
% music21 (key, timesig) mido (duration, tempo, instruments)
We use Music21 \cite{cuthbert2010music21} and Mido \cite{mido} libraries to extract the following features from MIDI: Musical Key (Music21), Time Signature (Music21), Tempo (Mido), Duration of the MIDI file (Mido), and a list of Instruments (Mido).

The \textbf{Key} and \textbf{Time Signature} features are obtained through \texttt{music21.midi.analyze('keys')} and \texttt{music21.midi.getTimeSignatures()} functions, respectively. To calculate the \textbf{Tempo}, we first look for the set\_tempo MIDI message to get the MIDI tempo. Then, the \texttt{mido.tempo2bpm()} function is used to convert this MIDI tempo to beats per minute (bpm). For MIDI file \textbf{Duration}, we retrieve the \texttt{length} attribute of a \texttt{mido.MidiFile} object.

% instruments
To extract a list of \textbf{instruments}, we filter the MIDI messages based on channel number and their associated instrument program obtained from the program change message. To treat ambiguity given by some faulty files, we always take the last assigned program number as the definite instrument number for each MIDI channel. For channel 10, which is reserved for drums, we always consider the assigned instrument to be drums, unless there is another percussion instrument specified.

We further process the extracted instruments in three steps to identify the most prominent instruments. First, we extract total note duration for each of the instruments by scraping through note-on and note-off messages, and rank them by this  duration. Second, we map the program numbers to their respective instrument names, grouping similar variations (e.g., both nylon and steel string acoustic guitars as `acoustic guitar') . Third, we reduce the list of instruments to only include one instance of the same instrument name (in the previous example, the two acoustic guitars would merge into one), and then take top five instruments sorted by their total note duration.

\subsubsection{Synthesized audio feature extraction}

We use the Midi2Audio library \cite{midi2audio} that utilizes FluidSynth \cite{fluidsynth,fluidsynth2} to synthesize audio from MIDI with the Fluid Release 3 General-MIDI sound font. Then, we use these audio files to extract genre, mood, and chord features.

To extract \textbf{genre} and \textbf{mood}, we use Essentia \cite{bogdanov2013essentia}, specifically the MTG-Jamendo genre and mood/theme discogs effnet models\footnote{\rurl{essentia.upf.edu/models.html\#discogs-effnet}}. We keep the top two genre tags with the highest confidence score, and the top five mood/theme tags, also based on their confidence score.  The confidence scores for each tag are also stored.

Next, we extract the single most occurring \textbf{chord sequence} of length 3 to 5. To obtain this, we first extract all chords from the audio using Chordino \cite{chordino}. To obtain the most frequent short chord sequence, 
% iterate through the list of ext
we first iterate through the chord list to find the most frequent patterns consisting of 3, 4, and 5 consequent chords. We do not allow these patterns to have the same first and last chord, e.g., [A, B, C, A] for a pattern of length 4 is not allowed, as this is likely an [A, B, C] pattern of length 3. Then, we decide on which pattern to keep through a set of rules described in Algorithm~\ref{alg:chords}.
In the below algorithm $n_i$ represents the occurrence count of the most frequent pattern ($p_i$) of length $i$. 
We save the final selected pattern along with a number denoting how many times it occurred. Once we have extracted all of the features extracted, we move on to caption generation, described in the next subsection.

% \vspace{-1mm}
\begin{algorithm}[h!]
        \caption{Selecting the most frequent chord pattern. }
        \label{alg:chords}
\begin{algorithmic}
\Parameter{$p_i$}{most frequent pattern of length $i$}
\Parameter{$n_i$}{occurrence count of $p_i$}
\Parameter{$p$}{final selected most frequent pattern}
\State $n=n_3+n_4+n_5$
\If{$(n_5\geq 0.8\cdot n_4) \And (n_5\geq 0.25\cdot n)$} 
    \State $p \gets p_{5}$
\ElsIf{$(n_4\geq 0.8\cdot n_3) \And (n_4\geq 0.3\cdot n$} 
    \State $p \gets p_{4}$
% \ElsIf{$(\text{\text{count}}_3=0) \And (\text{\text{count}}_4=0) \And (\text{\text{count}}_5=0)$} 
%     \State $p \gets None$
\ElsIf{$(n_3==0)$} 
    \If{$(n_4==0)$}
        \If{$(n_5==0)$}
            \State $p \gets \text{None}$
        \Else
            \State $p \gets  p_{5}$
        \EndIf
    \Else
        \State $p \gets  p_{4}$
    \EndIf
\Else
    \State $p \gets  p_{3}$
\EndIf
\end{algorithmic}
% \vspace{-4mm}
\end{algorithm}
% \vspace{-6mm}
% \vspace{-2 cm}

% $
% total\_count=\text{\text{count}}_3+\text{\text{count}}_4+\text{\text{count}}_5
% If (\text{\text{count}}_5>=0.8*\text{\text{count}}_4) & \text{\text{count}}_5>0.25*total\_count
%     -> Choose pattern of 5 as the final pattern
% If \text{\text{count}}_4)>=0.8*\text{\text{count}}_3
% etc.
% 

% \begin{algorithmic}
% \State $\text{total}_{\text{count}}=\text{\text{count}}_3+\text{\text{count}}_4+\text{\text{count}}_5$

% \If{$(\text{\text{count}}_5\geq 0.8\cdot \text{\text{count}}_4) \And (\text{\text{count}}_5\geq 0.25\cdot \text{total}_{\text{count}})$} 
%     \State $p_{\text{final}} \gets p_{5}$
% \ElsIf{$(\text{\text{count}}_4\geq 0.8\cdot \text{\text{count}}_3) \And (\text{\text{count}}_4\geq 0.3\cdot \text{total}_{\text{count}})$} 
%     \State $p_{\text{final}} \gets p_{4}$
% % \ElsIf{$(\text{\text{count}}_3=0) \And (\text{\text{count}}_4=0) \And (\text{\text{count}}_5=0)$} 
% %     \State $p_{\text{final}} \gets None$
% \ElsIf{$(\text{\text{count}}_3==0)$} 
%     \If{$(\text{\text{count}}_4==0)$}
%         \If{$(\text{\text{count}}_5==0)$}
%             \State $p_{\text{final}} \gets None$
%         \Else
%             \State $p_{\text{final}} \gets  p_{5}$
%         \EndIf
%     \Else
%         \State $p_{\text{final}} \gets  p_{4}$
%     \EndIf
% \Else
%     \State $p_{\text{final}} \gets  p_{3}$
% \EndIf
% \end{algorithmic}
\subsection{Caption generation}
\label{sec:caption}
In this step, we take the extracted features and execute caption generation. To harness the expressive power and few-shot capability of a Large Language Model (LLM), we refer to a recent benchmarking study on LLMs \cite{kevian2024capabilities}, and ultimately selected Claude 3 Opus \cite{claude3} due to its superior performance compared to other LLMs such as GPT4. Employing in-context learning (a task in which the LLM is given example data of paired input-output to serve as `context', and is expected to continue producing outputs for new unpaired inputs in a similar manner), we begin by selecting 17\footnote{Optimized based on limit on input tokens in Claude 3 text prompts.} diverse examples from the extracted features and request a human annotator to craft appropriate text captions for each of these based solely on the extracted features. This approach aims to prevent any auditory influence on human captioning, as Claude 3 (or any LLM, for that matter) will subsequently only process text inputs, not audio files. 
% Once the 17 examples are prepared, we construct a text prompt that instructs Claude 3 to analyze the human-prepared feature-caption pairs, and generate suitable captions for new sets of features given this context. To maintain clarity and conciseness, we specify in our prompt to limit the generated captions between three to seven sentences. Before performing caption generation for all 168K MIDI files, we conduct a sanity check on a subset of ten examples. Please note, this step evaluates Claude 3's response to in-context learning, allowing us to tweak our instruction prompt to ensure that it does not produce unrelated output or "hallucinate". This check differs from the quality evaluation of the generated captions reported in the next section. In our study, a single round of sanity checks sufficed, obviating the need to modify prompts or alter the feature-caption pairs for in-context learning. Finally, leveraging the features extracted from each MIDI file, we generate corresponding captions, thus culminating in the creation of our proposed \textbf{\dataset} dataset, which we describe in detail in the next section.
Once the 17 examples are prepared, we construct a text prompt instructing Claude 3 to analyze the human-prepared feature-caption pairs and generate suitable captions for new sets of features. To maintain clarity, we specify that the generated captions should be between three to seven sentences. Before generating captions for all 168K MIDI files, we conduct a sanity check on ten examples to evaluate Claude 3's response to in-context learning, ensuring our prompt does not produce unrelated output or "hallucinate." Please note, this check differs from the quality evaluation of the generated captions reported in the next section. In our study, a single round of sanity checks sufficed, obviating the need to modify prompts or alter the feature-caption pairs for in-context learning. Finally, using the features extracted from each MIDI file, we generate corresponding captions, creating our proposed \textbf{\dataset} dataset, which we describe in detail in the next section.
% Further details regarding the resulting dataset are provided in the subsequent section.
\begin{figure*}[h!]
\centering
\begin{subfigure}{0.49\textwidth}
  \centering
  \includegraphics[width=\linewidth]{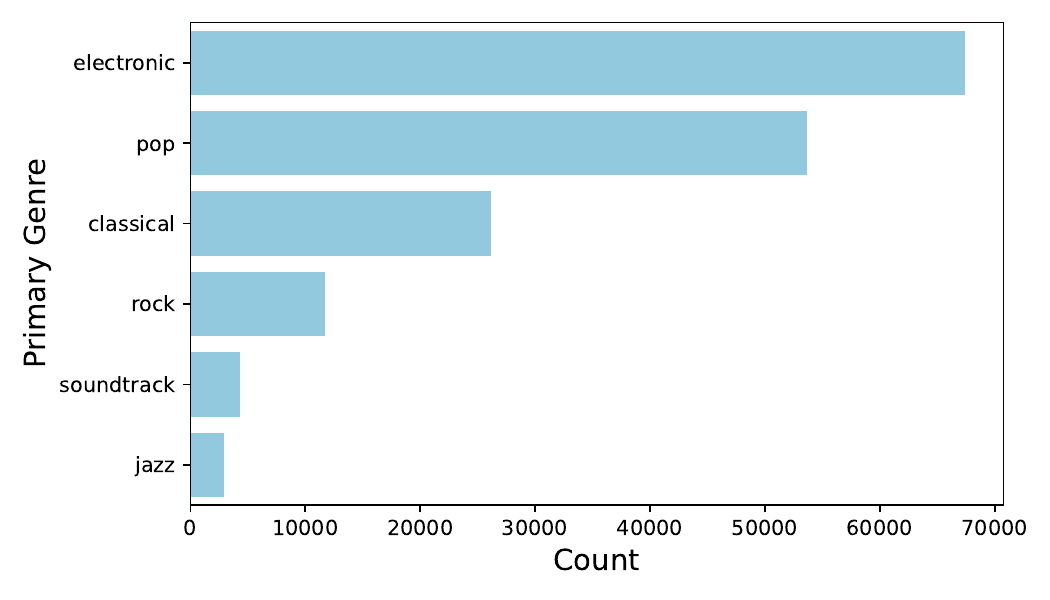}
  \caption{Distribution of primary genre.}
  \label{fig:genre}
\end{subfigure}
\hfill
\begin{subfigure}{0.49\textwidth}
  \centering
  \includegraphics[width=\linewidth]{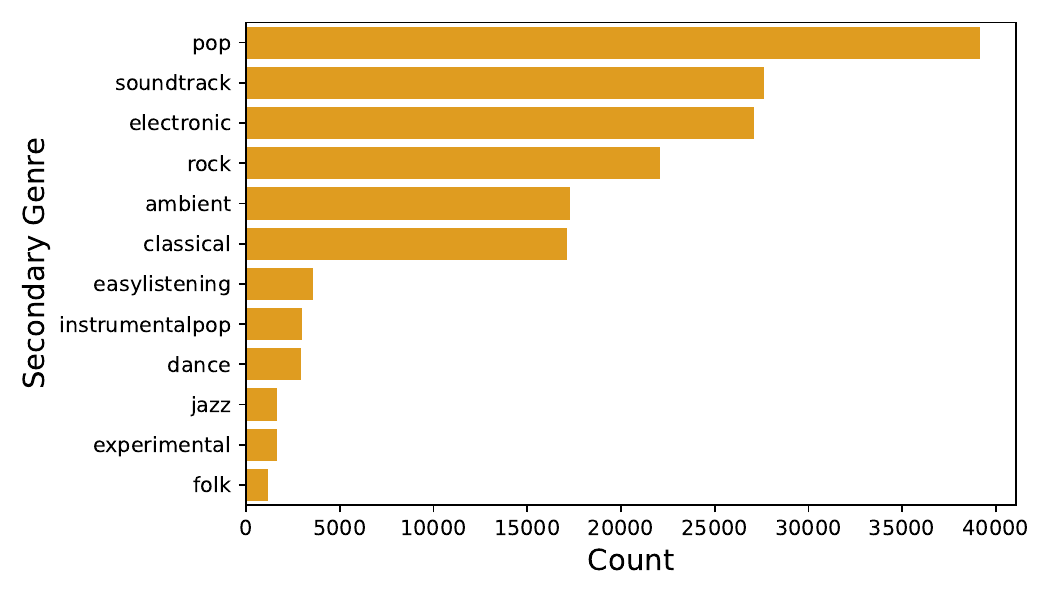}
  \caption{Distribution of secondary genre. }
  \label{fig:secondary_genre}
\end{subfigure}
\caption{Genre distributions. }
\vspace{-2mm}
\end{figure*}
\begin{figure*}[h!]
\centering
\begin{subfigure}{0.49\textwidth}
  \centering
  \includegraphics[width=\linewidth]{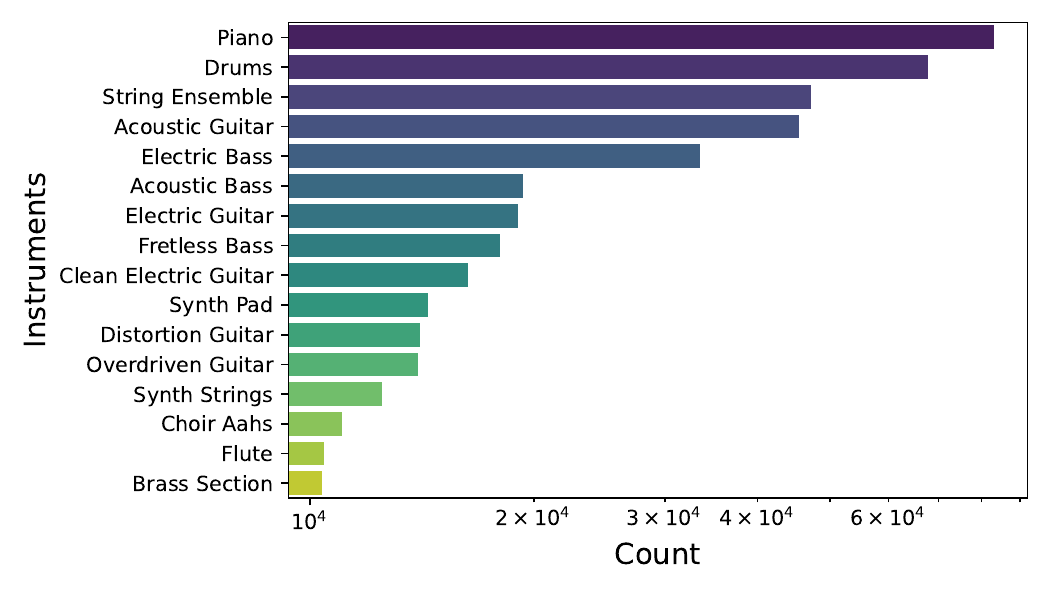}
  \caption{Distribution of instruments present in the summary.}
  \label{fig:instruments}
\end{subfigure}
\hfill
\begin{subfigure}{0.49\textwidth}
  \centering
  \includegraphics[width=\linewidth]{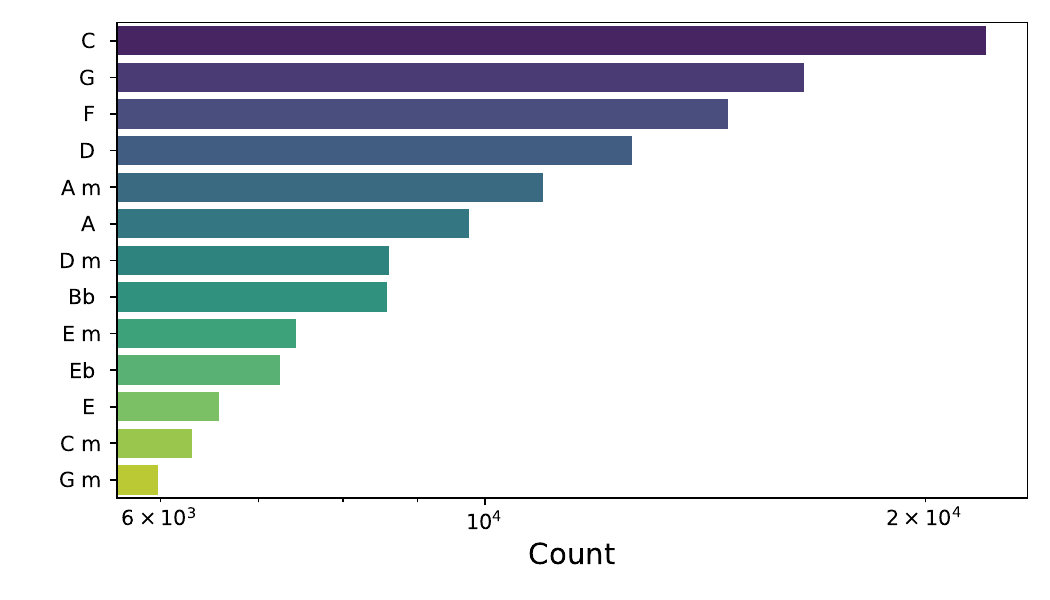}
  \caption{Distribution of key.}
  \label{fig:key}
\end{subfigure}
\caption{Instrument and key distributions (in log scale). }
\vspace{-2mm}
\label{fig:inskey}
\end{figure*}
\begin{figure*}[h!]
\centering
\begin{subfigure}{0.49\textwidth}
  \centering
  \includegraphics[width=\linewidth]{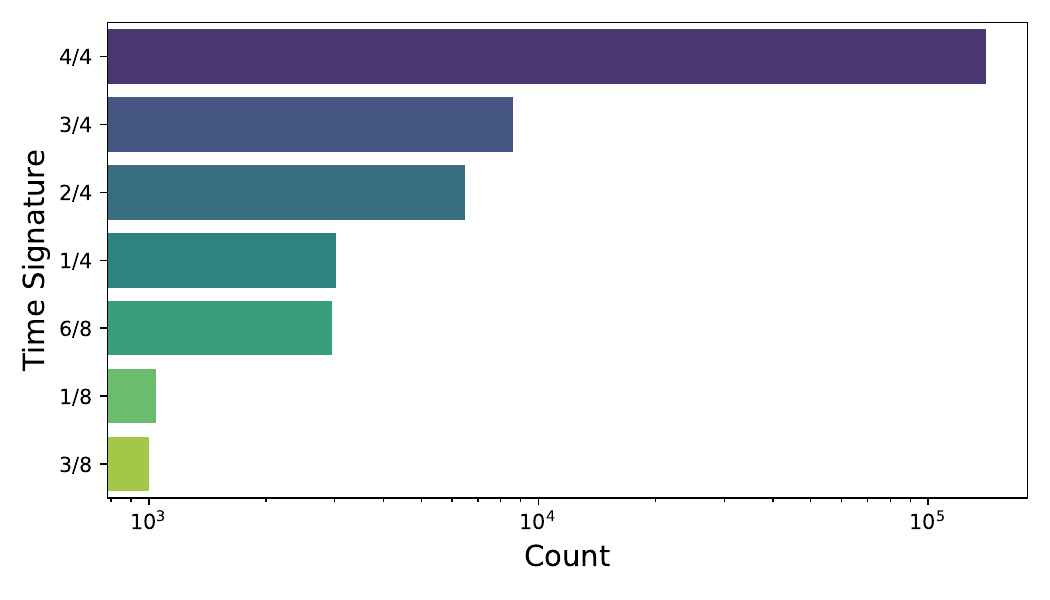}
  \caption{Distribution of time signature.}
  \label{fig:time_signature}
\end{subfigure}
\hfill
\begin{subfigure}{0.49\textwidth}
  \centering
  \includegraphics[width=\linewidth]{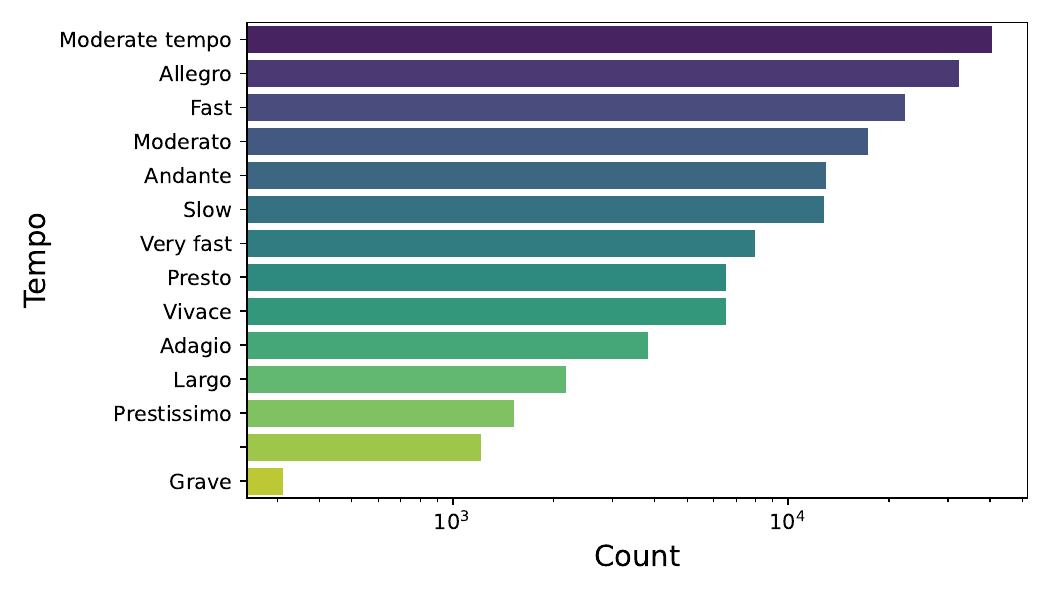}
  \caption{Distribution of tempo.}
  \label{fig:tempo}
\end{subfigure}
\caption{Time signature and tempo distributions (in log scale).}
\vspace{-4mm}
\end{figure*}
% \vspace{-0.2cm}
\section{Evaluation and statistics}
In this section, we first introduce the \textbf{\dataset} dataset and subsequently detail subjective evaluation in form of listening study.
\subsection{\textbf{\dataset} dataset}

% \dh{Introduce the original Lakh midi dataset first, provide reference, and talk about size etc. Then go into statistics. }

% \dh{Can we display 2 generated captions or so in a table? As an example? }
To generate our \textbf{\dataset} dataset, we start with MIDI files provided in the Lakh MIDI dataset \cite{raffel2016learning}, comprised of a collection of 176,581 unique MIDI files, designed to facilitate large-scale music information retrieval. Additionally, the dataset is distributed under a CC-BY 4.0 license, enabling us to expand the dataset without encountering copyright constraints. Subsequently, we process the raw MIDI files and extract musical features as described in Section \ref{sec:feat}, which we used in the captioning process {Section \ref{sec:caption}} to create our final \textbf{\dataset} dataset consisting of 168,407 MIDI files with matching text caption. A couple of examples of captions generated are provided below. They encapsulate key information regarding the music contents while infusing a fluid human touch:
\begin{enumerate}
% \item "A melodic and happy electronic pop song featuring distortion guitar, orchestral harp, slap bass, piano, and string ensemble. The song is in the key of F major with a 4/4 time signature and follows a chord progression of Bbmaj7, Gm, Am7, Dm7, and E7. It has a corporate and energetic vibe with a hint of darkness."
\item ``A melodic and happy rock and pop song featuring a string ensemble, piano, clean electric guitar, slap bass, and drums. The song is in the key of F major with a 4/4 time signature and a tempo of 120 BPM. The chord progression alternates between Bb and F throughout the song, creating a motivational and energetic corporate vibe.''

    \item ``A melodic and happy pop song with a Christmas vibe, featuring piano, clean electric guitar, acoustic guitar, and overdriven guitar. The song is in the key of A major with a 4/4 time signature and a moderate tempo. The chord progression revolves around D, E6, D, and E, creating a motivational and loving atmosphere throughout the piece.''
\end{enumerate}
\par
Moreover, we provide a summary of some of the extracted features below to gain further insight into the diversity within the dataset. In Figure~\ref{fig:genre}, we illustrate the distribution of the primary (highest confidence score) and secondary (second highest confidence score) genres present in the dataset. In both cases, electronic and pop genres are most prominent in the dataset. The secondary genre exhibits more variation, such as folk, instrumental pop, and easy listening, which have more occurrences as secondary genre but do not appear in the primary genre figure. This means that they can be used by the captioning system to further specify and narrow down the broad primary genres (e.g. classical) into more specific descriptions such as `ambient classical' etc. Please note that only genres with more than 1,000 occurrences are displayed in the figures. Figure~\ref{fig:inskey} summarizes the instruments and keys present in the dataset. Piano, drums, and various types of guitars are predominant in the instrument summary, corroborating the fact that a significant portion of the songs belongs to electronic, pop, and rock genres. Additionally, the keys of C, G, F, and D major have the highest occurrences in the dataset. Regarding time signature, 4/4 is significantly more common than any other (Figure~\ref{fig:time_signature}
), while most songs follow a moderate tempo (Figure~\ref{fig:tempo}). 

\subsection{Listening study}
\label{sec:list_study}
% There is no ground truth or baseline model to compare our new dataset with. Hence, with the help of PsyToolkit \cite{stoet2010psytoolkit,stoet2017psytoolkit}, we conduct two listening studies whereby listeners were asked to rate the quality of generated captions. In the first, general audience study, we asked participants to listen to up to 20 MIDI files, chosen at random, and rate the captions in terms of naturalness and accuracy. The results of this study, collected from 35 participants, are shown in Table~\ref{tab:results_general} and show that the captions are natural sounding and summarize the audio content well. We have kept or study open to participation, inviting and encouraging additional participants to evaluate our captions. We intend to update the table as we progress through later stages of submission and gather more responses. We observed that even a single mismatched word in the caption might lead to significant reductions in raters' scores due to the specific nature of our inquiries. Additionally, the inclusion of music-technical details like key and chords in the captions might cause confusion among the raters. Hence, we conducted a second experiment with two music experts who have absolute pitch. The results are shown in Table~\ref{tab:results_expert}. 

Since there is no ground truth or baseline model to compare our new dataset to, we conduct a listening study with the help of the PsyToolkit platform \cite{stoet2010psytoolkit,stoet2017psytoolkit}.
% There is no ground truth or baseline model to compare our new dataset to. Hence, with the help of the PsyToolkit platform \cite{stoet2010psytoolkit,stoet2017psytoolkit}, we conduct a listening study.
Listeners were asked to listen to 20 MIDI files, chosen at random, from which 15 are captioned by our framework and 5 are annotated by an expert human rater with absolute pitch. Then, listeners were asked to rate these captions in seven aspects, which are: 1) Overall matching of caption to audio, 2) How human-like the caption is, 3) Genre matching of caption with audio, 4) Mood matching, 5) Key matching, 6) Chord matching, and 7) Tempo matching. Those listeners who indicated that they do not have the ability to recognize chords/key were tagged as General audience. A total of 16 participants belong to this general audience, of which 25\% has more than 1 year of musical training. Another 7 participants indicated that they can recognize chords and key or have absolute pitch. These were tagged as Music experts.

\subsection{Results and discussion}
\label{sec:res_dis}

Table~\ref{tab:results_general} shows the results of the listening study. The average rating for overall matching of the text caption with the MIDI file for the general audience is even slightly higher (5.63) for the AI generated caption compared to the human-written caption (5.46). When it comes to the ratings by music experts, the overall matching rating is slightly lower, but still well above average (4.92). In term of how human-like the captions are, the general audience again provides high ratings, comparable to those given to the human-written captions (5.21). The music experts are slightly more critical and rate them at 4.98, which is still very close to their rating for human-written captions (5.09). A similar pattern can be seen for ratings of genre matching and mood matching. 
The ratings for tempo matching outperformed the human-written ones for both general audience and music experts.  

\begin{table}[t] 
    \centering
    \begin{tabular}{l|cc|cc}
    \toprule
        \textbf{Audience:} & \multicolumn{2}{c}{\textbf{General audience}} & \multicolumn{2}{|c}{\textbf{Music experts}}  \\ 
        % & Human-annotated & AI-annotated & Human-annotated & AI-annotated \\
        \textbf{Annotated by:} & \textbf{Human} & \textbf{AI} & \textbf{Human} & \textbf{AI} \\
        \midrule
         % Question & Score (1-7) & Score (1-7) & Score (1-7) & Score (1-7)\\
        Question & \multicolumn{4}{c}{Avg. rating (1-7)} \\
         \midrule
        Overall matching & 5.46 & 5.63 & 5.40 & 4.92  \\
        Human-like & 5.21 & 5.32 & 5.09 & 4.98 \\
        Genre matching & 5.80 & 5.63 & 5.54 & 4.73 \\
        Mood matching & 5.50 & 5.62 & 5.43 & 4.82 \\
        Key matching & 5.87 & 5.70 & 5.51 & 5.69 \\
        Chord matching & 6.12 & 5.78 & 5.74 & 5.09 \\
        Tempo matching & 5.71 & 5.86 & 5.37 & 5.77 \\
         \bottomrule
    \end{tabular}
    \caption{Results of the listening study. Each question is rated on a Likert scale from 1 (very bad) to 7 (very good). The table shows the average ratings per question for each group of participants. }
    \label{tab:results_general}
    \label{tab:results_expert}
    \vspace{-4mm}
\end{table}

In terms of key and chord matching, the general audience provide good ratings. For these questions the ratings from the music experts, however, are more reliable, as these participants have explicitly indicated that they are able to recognize chords and keys. Their rating for key matching (5.69) is on par with the rating for human-written captions (5.51), and confirm the high agreement that the musical key described in the caption matches the audio pieces. For chord matching, the music experts' average rating of 5.09 falls below the rating for the human-written caption. Please note, however, that this particular question was not easy to answer.
% The task of extracting a single `main' pattern (3-5 chords) from the entire list of extracted chords, to be used in a text caption is quite challenging as there are many different cases, e.g., very short fragments of a few chords, very long pieces with many chord patterns inside, slow songs with a low number of chord changes, or pieces with many simultaneous instruments playing, which can cause slight variations in the extracted chords. Slight changes in chord patterns can also be intentional, e.g., a chord progression of C, G, D, C, G, D6 would likely be detected as a C, G, D, C, G pattern instead of a C, G, D variation.
Extracting a single `main' pattern (3-5 chords) from the entire list of extracted chords is challenging as there are many different cases, e.g., very short fragments of a few chords, and very long pieces with many chord patterns. Slight changes in chord patterns can also be intentional, e.g., a chord progression of C, G, D, C, G, D6 would likely be detected as a C, G, D, C, G pattern instead of a C, G, D variation.
All this makes it hard to objectively judge a single-chord pattern in the text captions. Despite this, the chord matching rating of 5.09 provides support that our caption contains a matching chord summary. Overall, the results from the listening study support that our text captions provide a high-quality, human-like textual description that matches the MIDI files well.

The task of automatically labelling files of various length is difficult by nature as longer music pieces might require more text to be described precisely, while shorter pieces may need only a single sentence. This problem is further magnified when considering chord progressions and their summary as mentioned above. 
Additionally, extracting features from synthesized audio files is not optimal, as the choice of the sound font has an impact on the obtained results, which is likely to be most apparent in genre and mood features. Future research could focus on improving accuracy related to these features.
In sum, we are confident that our \textbf{\dataset} dataset will facilitate the development of the first Text-to-MIDI generation models.
%In addition, it is very hard and psychologically tiring for humans to transcribe chords. Finally, even for an AI algorithms, no perfect chord detection model currently exists. 

%The expert listening test confirms that the captions match the audio clips well and that they are natural sounding (rating of 6.1 and 6.8, respectively). 

%There is also a high agreement that the musical key described in the caption matches the audio pieces (rating 6.2). 

%In terms of chord matching, we see a decrease in rating, which is likely caused by the chords of full-length songs being reduced to a main chord sequence containing only a few chords.
% \vspace{-5mm}
\section{Conclusion}
We present the first large-scale open MIDI captioned dataset, \textbf{\dataset}.  This dataset also includes a comprehensive set of musical features such as chord patterns, genre, and mood. To facilitate the development of this dataset, we have developed a MIDI captioning framework. This approach includes music feature extraction and summarization from MIDI and the synthesized audio, as well as the use of the Claude-3 LLM to generate the final captions using in-context learning. 
To evaluate the final dataset, we have conducted two subjective listening studies, which confirm that the captions are natural and indeed contain a text description of the musical features contained in the accompanying MIDI file. 
The resulting new \textbf{\dataset} dataset contains 168,407 MIDI files with descriptive text captions and is available online\footnote{\rurl{huggingface.co/datasets/amaai-lab/MidiCaps}} under a Creative Commons licence.

% contributions for reference: 
%  We introduce the first curated large-scale open dataset of MIDI-caption pairs, termed \textbf{MIDICaps}\footnote{Download url suppressed for anonymous review. }. Leveraging the in-context learning capability of large language models (LLMs), we enable the generation of captions using only a small number of feature-caption training pairs.

% Furthermore, we present a comprehensive set of music-specific features extracted from MIDI files. These features succinctly characterize the musical content, encompassing tempo, chord progression, time signature, instrument presence, genre, and mood.

% Finally, we provide a text caption annotation framework tailored specifically for MIDI data. This framework, a first of its kind, is made freely accessible to users\footnote{Link suppressed for anonymous review. }, facilitating the generation of MIDI-caption pairs for their individual MIDI files.

\section{Acknowledgments}
This project has received funding from the SUTD Kickstarter Initiative no. SKI 2021\_04\_06. 

% DH: later on, to acknowledge SKI funding

% For bibtex users:
\bibliography{main}

\end{document}